\documentclass[aps,12pt,superscriptaddress,amsfonts,amssymb,amsmath]{revtex4}
\pdfoutput=1

\usepackage{graphicx}
\usepackage{epsfig}
\usepackage{makeidx}
\usepackage{epstopdf}
\usepackage{natbib}
\usepackage{xcolor}

\usepackage{color}

\begin{document}


\title{Diagonal degree correlations vs.\ epidemic threshold \\ in scale-free networks}

\author{M.L.\ Bertotti \footnote{Email address: marialetizia.bertotti@unibz.it}}
\author{G.\ Modanese \footnote{Email address: giovanni.modanese@unibz.it}}
\affiliation{Free University of Bozen-Bolzano \\ Faculty of Science and Technology \\ I-39100 Bolzano, Italy}

\linespread{0.9}

\begin{abstract}
We prove that the presence of a diagonal assortative degree correlation, even if small, has the effect of dramatically lowering the epidemic threshold of large scale-free networks. The correlation matrix considered is $P(h|k)=(1-r)P^U_{hk}+r\delta_{hk}$, where $P^U$ is uncorrelated and $r$ (the Newman assortativity coefficient) can be very small. The effect is uniform in the scale exponent $\gamma$, if the network size is measured by the largest degree $n$. We also prove that it is possible to construct, via the Porto-Weber method, correlation matrices which have the same $k_{nn}$ as the $P(h|k)$ above, but very different elements and spectrum, and thus lead to different epidemic diffusion and threshold. Moreover, we study a subset of the admissible transformations of the form $P(h|k) \to P(h|k)+\Phi(h,k)$ with $\Phi(h,k)$ depending on a parameter which leave $k_{nn}$ invariant. Such transformations affect in general the epidemic threshold. We find however that this does not happen when they act between networks with constant $k_{nn}$, i.e. networks in which the average neighbor degree is independent from the degree itself (a wider class than that of strictly uncorrelated networks).

\end{abstract}

\maketitle

\section{Introduction}

From the mathematical point of view, a network is completely characterized (up to isomorphisms corresponding to simple re-denominations of the vertices) when a list of links or an adjacency matrix are given. In many applications involving large networks, however, one often summarizes the information on the network structure in a statistic-probabilistic form, by introducing the two fundamental quantities $P(k)$ and $P(h|k)$. $P(k)$, called degree distribution, represents the probability that a randomly chosen vertex of the network has $k$ neighbors. $P(h|k)$, called degree correlation function, expresses the conditional probability that a vertex of degree $k$ is connected to a vertex of degree $h$. An alternative but equivalent description involves the symmetric quantities $e_{jk}$, defined as the probabilities that a randomly chosen link connects two vertices of degree $j$ and $k$ \cite{boccaletti2006complex,newman2010networks,barabasi2016network}.

When one considers in a purely axiomatic way a class of networks, called Markovian networks \cite{boguna2003epidemic}, 
which are completely defined by assigning the quantities $P(k)$ and $P(h|k)$, 
one disregards higher-order correlations like e.g.\ $P(j|h|k)$ etc., which can in general be present. It is interesting to investigate the connections between real networks and the corresponding Markovian networks. In the case of Barabasi-Albert networks, for instance, it is possible to use recipes for constructing ensembles of the two kinds (preferential attachment vs.\ rewiring) and compare them \cite{bertotti2019configuration}.

In any case, let us focus on $P(k)$ and $P(h|k)$.
Imagine that we know them for a certain network and we want to study some dynamical processes on the network, e.g.\ epidemic diffusion processes \cite{pastor2015epidemic}. It turns out \cite{boguna2003absence,boguna2002epidemic} that several features of these processes depend on a ``contracted'' form of the correlations, namely the function
$k_{nn}(k)$, called ``average nearest neighbor degree'' and defined as $k_{nn}(k)=\sum_{k=1}^n hP(h|k)$, where $n$ is the highest degree of the nodes of the network. This function  of $k$ is simpler to analyse than the full matrix $P(h|k)$. Its increasing or decreasing character discriminates between assortative and disassortative networks (\cite{newman2003mixing,noldus2015assortativity,bertotti2020network,bertotti2021comparison} and refs.). We can further contract the information on the correlations into a single number, the Newman assortativity coefficient $r$, either using the the $e_{jk}$ matrix or with one more summation procedure performed on the $k_{nn}$.

One may wonder whether it is possible, given an admissible $k_{nn}$ function (it must satisfy a normalization condition, see below), to compute a full correlation matrix which returns that $k_{nn}$ upon contraction on $h$. Porto and Weber have devised a method for this purpose  \cite{weber2007generation}, which has been used for some applications by themselves and Silva et al.\ \cite{silva2019spectral}. However, while the correspondence $P(h|k) \to k_{nn}(k)$ is univocal, this is not true for the opposite correspondence $k_{nn}(k) \to P(h|k)$. One first scope of this work is to show explicitly this ambiguity in an important specific example, namely that of a linear $k_{nn}$. To this end we introduce in Sect.\ \ref{2-a} the correlation matrix $P^{Vaz-Wei}(h|k)$ of Vazquez-Weigt \cite{vazquez2003computational}, which has the simple form $(1-r)P^U_{hk}+r\delta_{hk}$, where $P^U$ is an uncorrelated matrix. The corresponding $k_{nn}$ is linear in $k$. Then, in Sect.\ \ref{2-b} we recall the method by Porto and Weber for building a $P(h|k)$ starting from a $k_{nn}$, and  in Sect.\ \ref{2-c} we apply it to the $k_{nn}$ of Vazquez and Weigt. A comparison of the result with the original matrix $P^{Vaz-Wei}(h|k)$ shows remarkable differences.

While examining these differences we have been led to consider the eigenvalue spectra of the associated connectivity matrices $C_{kh}=kP(h|k)$. This has revealed a simple general property of the eigenvalues of $C^{Vaz-Wei}$, which has important consequences for the epidemic threshold in diffusion models based on this matrix (Sect.\ \ref{differ}). In fact, the eigenvalues of $C^{Vaz-Wei}$ are $\Lambda^{(i)}=ri$, where $i=1,\ldots,n$ and $n$ is the largest degree in the network. It follows that the epidemic threshold $\lambda_c=1/\Lambda^{max}$ (Sect.\ \ref{connect}) is proportional to $n^{-1}$, for any fixed value of $r$. When $r$ is small, the epidemic threshold $\lambda_c$ is definitely greater than zero for small networks, but if $n \to \infty$ the threshold goes quickly to zero. The convergence is much faster than for other correlations, for which the largest eigenvalue typically grows as a root of $n$ or even as $\ln n$ when the scale-free exponent $\gamma$ is equal to 3 (Sects.\  \ref{uncorrel}, \ref{epidVW}). The conclusion is that adding even a very small amount of assortative diagonal degree correlations to an uncorrelated network leads, in the large-$n$ limit, to a fast vanishing of the epidemic threshold.

In Sect.\ \ref{4 Examples} we discuss a family of transformations of the correlation matrices which keep their $k_{nn}$ functions unchanged.
We show that such transformations affect in general the epidemic threshold, even if this does not happen when these transformations act on networks
in which the average neighbor degree is independent from the degree itself: in this case, the transformations lead to networks
for which the epidemic threshold remains unchanged
and which, albeit having a constant $k_{nn}$, belong to a wider class than that of strictly uncorrelated networks.

\section{The correlation matrix of Vazquez-Weigt vs.\ its Porto-Weber reconstruction}
\label{sec2}

\subsection{The Vazquez-Weigt matrix}
\label{2-a}

Vazquez and Weigt \cite{vazquez2003computational} have defined the following assortative correlation matrix:
\begin{equation}
P^{Vaz-Wei}(h|k)=(1-r)\frac{hP(h)}{\langle k \rangle}+r\delta_{hk} \, ,
\label{PVaz}
\end{equation}
where $\langle k \rangle=\sum_{k=1}^n k P(k)$. (More generally, below, $\langle g(k) \rangle$ denotes $\sum_{k=1}^n g(k) P(k)$ for any function $g$.)

This Ansatz has been used in several applications \cite{nekovee2007theory}. It is a linear combination of a perfectly uncorrelated matrix with elements $hP(h)/\langle k \rangle$, giving a probability of connection independent from $k$, and a perfectly assortative matrix $\delta_{hk}$ (giving a nonzero probability of connection only between nodes of the same degree). The coefficient $r$ in the linear combination can vary in the range $[0,1]$ and corresponds to the Newman assortativity coefficient.

The $k_{nn}$ function for the Vazquez-Weigt correlation matrix is easily found:
\begin{equation}
k^{Vaz-Wei}_{nn}(k)=\sum_{k=1}^n hP^{Vaz-Wei}(h|k)=(1-r)\frac{\langle k^2 \rangle}{\langle k \rangle}+rk \, .
\label{knnVaz}
\end{equation}
Since the first term is independent from $k$, this is a linear function, with slope $r$.

\subsection{The recipe of Porto-Weber for a correlation matrix having a pre-defined $k_{nn}$}
\label{2-b}

In order to compute the correlation matrix $P(h|k)$ starting from a given function $k_{nn}$ (normalized as $\sum_k kP(k)k_{nn}(k)=\langle k^2 \rangle$ \cite{bertotti2021comparison}), Porto and Weber define first the symmetric function
\begin{equation}
f(h,k)=1+\frac{[k_{nn}(h)-k_{me}][k_{nn}(k)-k_{me}]}{\langle k \, k_{nn}\rangle_e-k_{me}^2} \, ,
\label{fPW}
\end{equation}
where 
\[ 
k_{me}=\frac{\langle k^2 \rangle}{\langle k \rangle}
\]
and
\[
\langle k \, k_{nn}\rangle_e=\sum_h \frac{hP(h)}{\langle k \rangle} h k_{nn}(h) \, .
\]
In other words, $\langle k \, k_{nn}\rangle_e$ is the average of the quantity $k \, k_{nn}(k)$ with a normalized ``edge'' probability distribution defined as $P_e(k)=kP(k)/\langle k \rangle$, giving the probability that a randomly chosen edge of the network is connected to a node of degree $k$.

The conditional probability $P(h|k)$ is then given by
\begin{equation}
P(h|k)=\frac{hP(h)}{\langle k \rangle} \, f(h,k)
\label{PPW}
\end{equation}
It is immediate to check that $P(h|k)$ defined in this way satisfies the normalization condition in $h$ and the network closure condition
\[
h P(k|h) P(h) = k P(h|k) P(k), \, \qquad \hbox{for all} \ h, k  = i=1,...,n \, .
\]
Also it is straightforward to replace $P(h|k)$ into the definition of $k_{nn}(k)$ and obtain an identity.

\subsection{Porto-Weber recipe applied to the Vazquez correlation matrix}
\label{2-c}

Now, suppose we want to reconstruct $P^{Vaz-Wei}(h|k)$ starting from $k^{Vaz-Wei}_{nn}(k)$ using the Porto-Weber recipe.
Applying this recipe to the $k_{nn}$ function of Vazquez-Weigt
one obtains
\begin{equation}
    f(h,k)=1+r\frac{(h-k_{me})(k-k_{me})}{\langle k^3 \rangle /\langle k \rangle-k_{me}^2} \, .
\end{equation}

\begin{figure}[h]
  \begin{center}
\includegraphics[width=7.0cm,height=5.1cm]{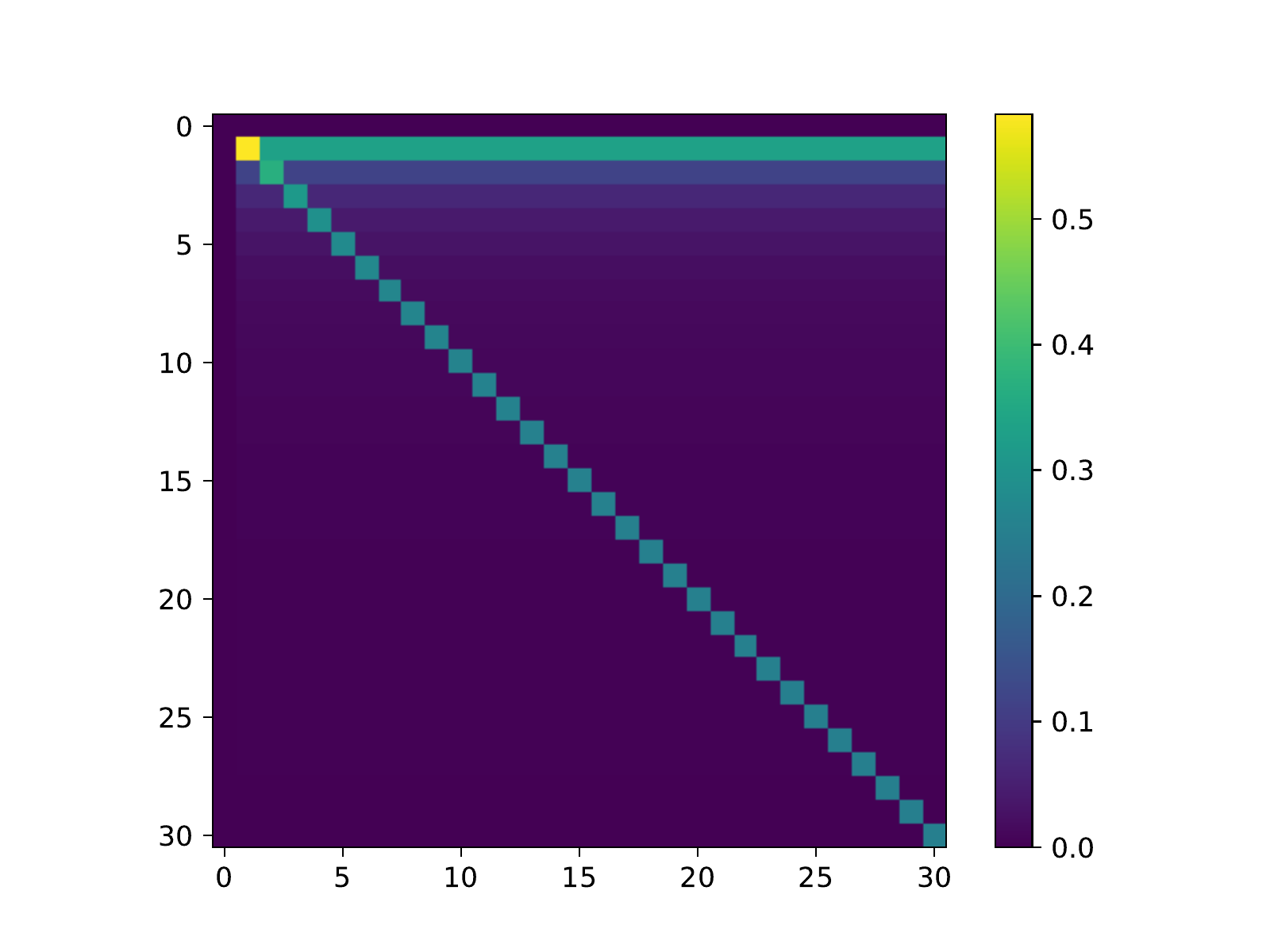}
\caption{The correlation matrix $P^{Vaz-Wei}(h|k)$ with $n=30$ (maximum degree), $r=0.25$ (Newman assortativity coefficient), $\gamma=2.5$ (scale-free exponent). The diagonal elements are clearly visible. Like for any correlation matrix, each column is normalized to 1. (Please note that row 0 and column 0 appear in the plot but do not actually belong to the matrix.)
} 
\label{PVaz-fig}
  \end{center}
\end{figure}

\begin{figure}[h]
  \begin{center}
\includegraphics[width=7.0cm,height=5.1cm]{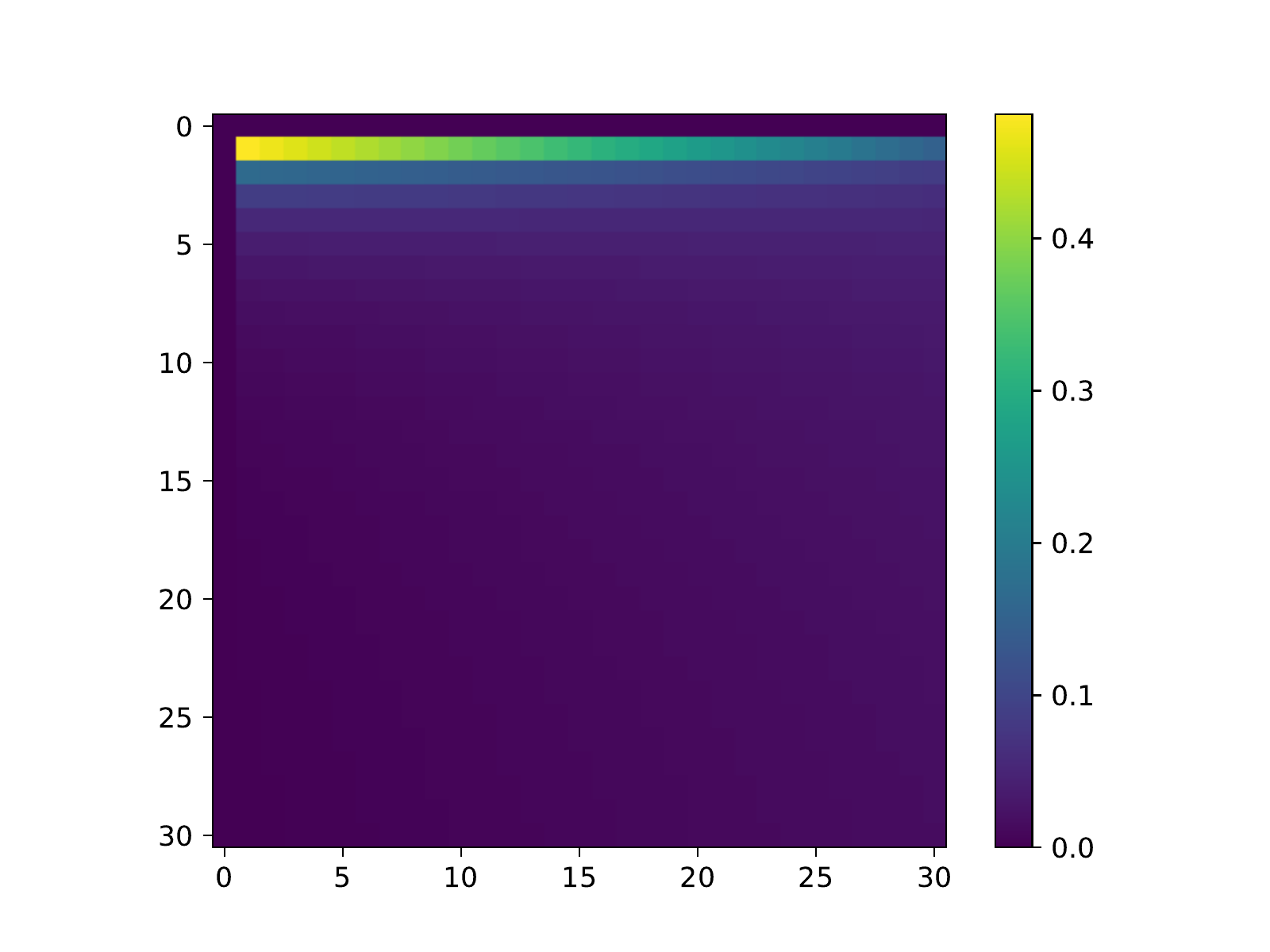}
\caption{The correlation matrix $P^{PW}(h|k)$ with $n=30$, $r=0.25$, $\gamma=2.5$. Note the difference with $P^{Vaz-Wei}(h|k)$ in Fig.\ \ref{PVaz-fig}, in spite of the fact that the two matrices have the same (linear) $k_{nn}$ function. In the ``deep blue'' lower part of the matrix there are variations which are not visible in this plot but are obviously needed to keep each column normalized to 1. Compare the example in Fig.\ \ref{riga25}.
} 
\label{PPW-fig}
  \end{center}
\end{figure}

One can check numerically that the insertion of this function $f(h,k)$ into the Porto-Weber recipe gives a correlation matrix $P^{Por-Web}(h|k)$ whose $k_{nn}$ coincides element by element with the $k_{nn}$ of Vazquez-Weigt. However $P^{Por-Web}(h|k)$ does not coincide with $P^{Vaz-Wei}(h|k)$. 
The difference between the two matrices is evident looking at their dependence on $h$ and $k$. Their traces and eigenvalues are markedly different, as we shall show in the next section. A graphical representation showing the differences of the single elements is given in Figs.\ \ref{PVaz-fig}, \ref{PPW-fig}.

In conclusion, with the Porto-Weber recipe it is possible to obtain from a given  $k_{nn}(k)$ function a correlation matrix $P(h|k)$ which yields that $k_{nn}$, but such a correlation matrix is not 
the unique correlation matrix having the given $k_{nn}$ as its average nearest neighbor degree function.
This could and should in fact be expected, since the definition of $k_{nn}$ involves a summation, and thus any two matrices $P^{(1)}(h|k)$ and $P^{(2)}(h|k)$, suitably normalized, such that
\begin{equation}
    \sum_{h=1}^n h [P^{(1)}(h|k)-  P^{(2)}(h|k)]=0
\end{equation}
yield the same $k_{nn}$.

\begin{figure}[h]
  \begin{center}
\includegraphics[width=10.5cm,height=7.6cm]{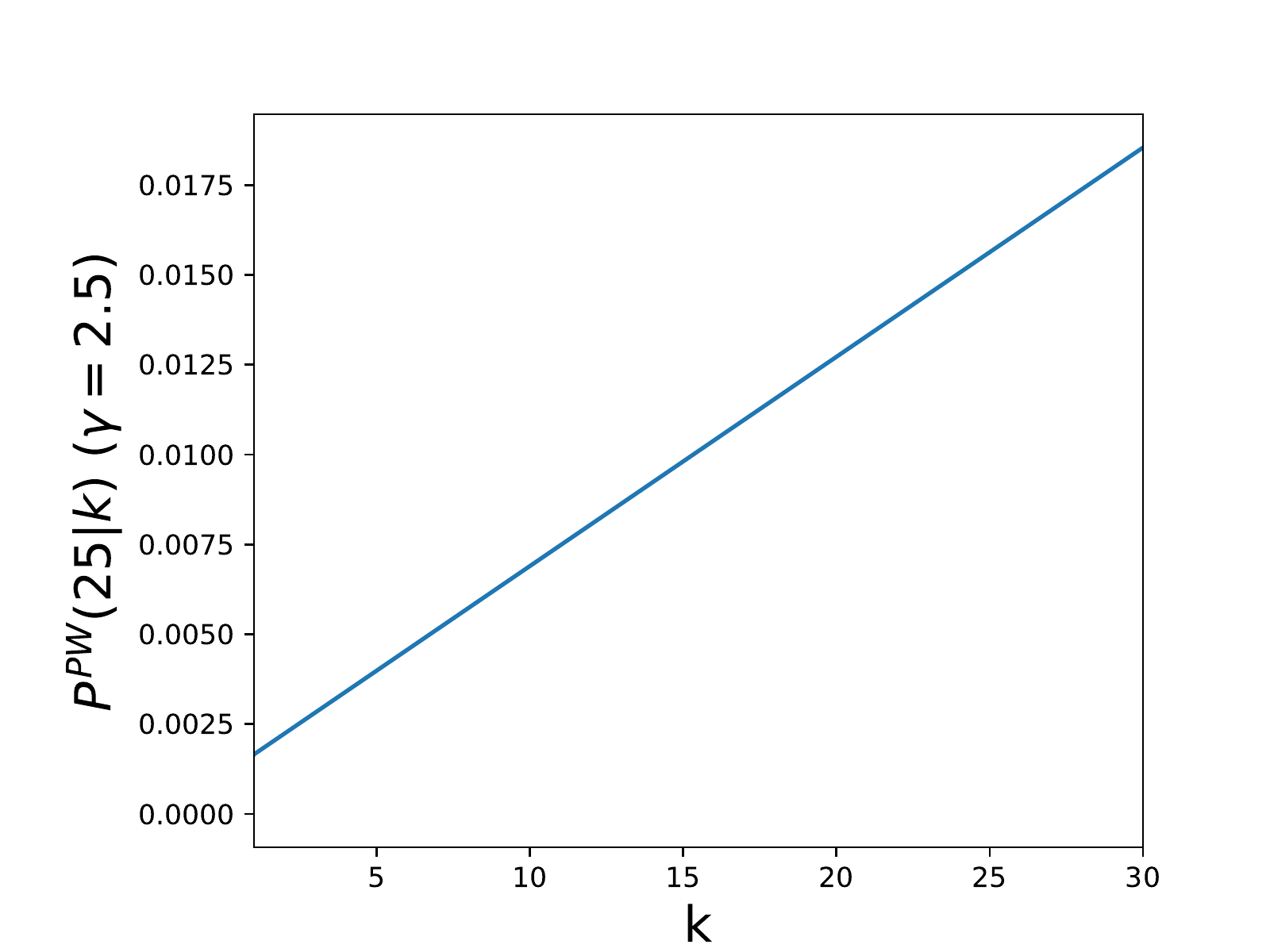}
\caption{The elements of row 25 of the matrix $P^{Por-Web}$ of Fig.\ \ref{PPW-fig}. There is an increase in $k$ which is not visible in the color scale of Fig.\ \ref{PPW-fig} but is necessary to preserve the normalization of the columns. The same happens of course for the other rows.
} 
\label{riga25}
  \end{center}
\end{figure}

We also observe that there is no guarantee that the Porto-Weber method works for \emph{any} normalized $k_{nn}$. For instance, for a linear $k_{nn}$ some (unacceptable) negative values of $P^{Por-Web}(h|k)$ are obtained when $r$ is greater than a value which is approximately  $0.5$. For the $k_{nn}$ functions of Ref.\ \cite{silva2019spectral}, of the form $k_{nn}=ck^\alpha$, one obtains negative values of $P^{Por-Web}(h|k)$ when $\alpha$ is greater than a value which is approximately $0.4$. 

\section{Differences in the spectrum of the connectivity matrix}
\label{differ}

\subsection{The connectivity matrix and its relation with the epidemic threshold}
\label{connect}

For a Markovian network with correlation matrix $P(h|k)$ the associated ``connectivity matrix'' is defined as \cite{boguna2003absence,boguna2002epidemic,silva2019spectral}
\begin{equation}
    C_{kh}=kP(h|k) \, .
\end{equation}

This matrix plays an important role in studies of diffusion on networks. For instance, in the Homogeneous Mean Field approximation of the SI (Susceptible-Infected) epidemic model, the equation set which describes the behavior in time of the fraction $\rho_k$ of infected nodes with degree $k$ is (see \cite{boguna2002epidemic})
\begin{equation}
    \frac{d\rho_k}{dt}=-\rho_k + (1-\rho_k) \lambda \sum_{h=1}^n kP(h|k)\rho_h \, ,   \qquad k=1,\ldots,n \, .
\end{equation}

It can be generally shown that the solutions of this equation set are characterized by an ``epidemic threshold'' $\lambda_c$ which separates different spreading scenarios:
if $\lambda>\lambda_c$, the system reaches a stationary state with a finite fraction of infected population, while if $\lambda<\lambda_c$, the contagion dies out exponentially fast. The threshold $\lambda_c$ turns out to be equal to $1/\Lambda^{max}$, where $\Lambda^{max}$ is the largest eigenvalue of the connectivity matrix $C$.

\subsection{The epidemic threshold for uncorrelated scale-free networks}
\label{uncorrel}

It is therefore important to know the largest eigenvalue of the $C$ matrix, and a general result valid for scale-free networks \cite{boguna2003absence,boguna2002epidemic} states that this eigenvalue tends to $+\infty$ when the size of the network grows. This means that for large scale-free networks the epidemic threshold is essentially zero and the epidemics spreads and persists in the population also when the contagion probability is very small. 

In the absence of degree correlations (i.e., with uncorrelated $C$, namely $C_{kh}^U=kP^U(h|k)=khP(h)/\langle k \rangle$), it has been shown that
\begin{equation}
    \Lambda^{max}=\frac{\langle k^2 \rangle}{\langle k \rangle} \, .
\end{equation}
For scale-free networks with scale exponent $2<\gamma<3$, $\langle k \rangle$ is finite when the maximum degree $n$ tends to infinity, while $\langle k^2 \rangle$ is divergent:
\begin{equation}
    \langle k^2 \rangle \simeq \int_{k_{min}}^n \frac{c_\gamma}{k^\gamma} k^2 dk = \left[ \frac{k^{3-\gamma}}{3-\gamma} \right]_{k_{min}}^n \, ,
    \label{k2}
\end{equation}
where $c_\gamma$ is the normalization constant of the degree distribution $P(k)=c_\gamma/k^\gamma$. The divergent part of (\ref{k2}) is of the form $n^{3-\gamma}$, thus $\lim_{n\to \infty} \Lambda^{max}=+\infty$ for $\gamma \in [2,3)$. When $\gamma=3$ the limit is also infinite, but only with slow divergence $\sim \ln n$.

\subsection{The epidemic threshold for the assortative networks of Vazquez-Weigt}
\label{epidVW}

It can be shown through general arguments that the divergence of the largest eigenvalue $\Lambda^{max}$ when $n\to \infty$ holds true independently from the degree correlations  (\cite{boguna2003absence,boguna2003epidemic}; see also some special cases in \cite{bertotti2021comparison}). However in the case of the diagonal assortative correlation matrices introduced by Vazquez and Weigt a simple direct proof is possible, which is not yet available in the literature. Since for these correlations the assortativity level (expressed through the Newman coefficient $r$) is easily tunable in the range $[0,1]$, the proof also has interesting consequences for the epidemic threshold in general.

\begin{figure}[h]
  \begin{center}
\includegraphics[width=10.5cm,height=7.6cm]{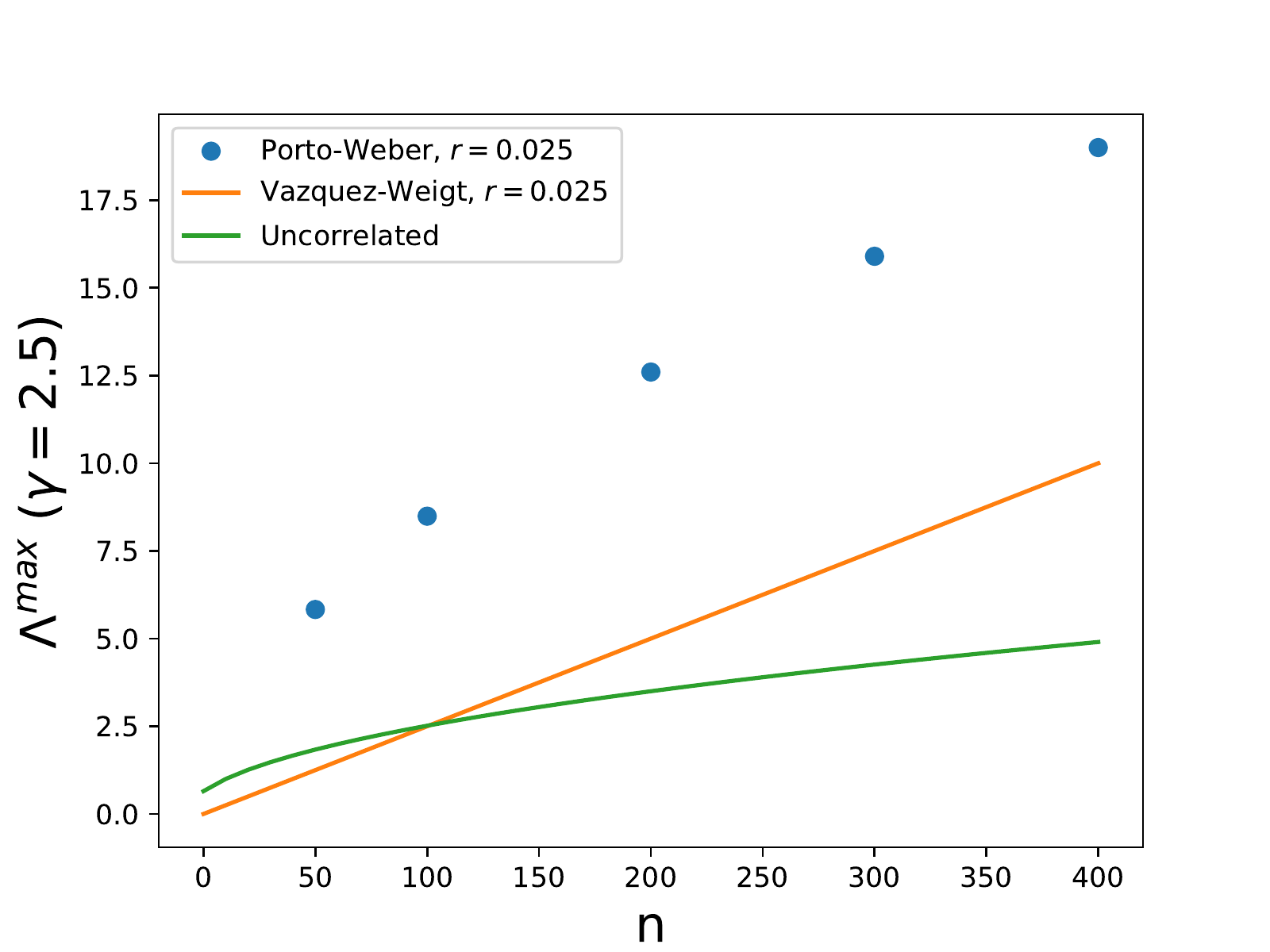}
\caption{Behavior, as a function of the maximum degree $n$, of the largest eigenvalue $\Lambda^{max}$ of the connectivity matrix, with scale-free exponent $\gamma=2.5$. The two continuous lines represent the case of uncorrelated networks and networks with diagonal correlations of the Vazquez-Weigt kind and assortativity coefficient $r=0.025$. The dots represent the case of networks reconstructed with the Porto-Weber recipe from the $k_{nn}$ function of the Vazquez-Weigt networks. Note that when $n$ increases the largest eigenvalue increases faster (linearly) for the Vazquez-Weigt networks than for uncorrelated networks, even though the assortativity is very small. Also for the Porto-Weber networks the increase is faster than for uncorrelated networks, but it is less than linear, and this gets worse when $\gamma$ increases (compare Figs.\ \ref{La-max-275}, \ref{La-max-3}).
} 
\label{La-max-25}
  \end{center}
\end{figure}

\begin{figure}[h]
  \begin{center}
\includegraphics[width=10.5cm,height=7.6cm]{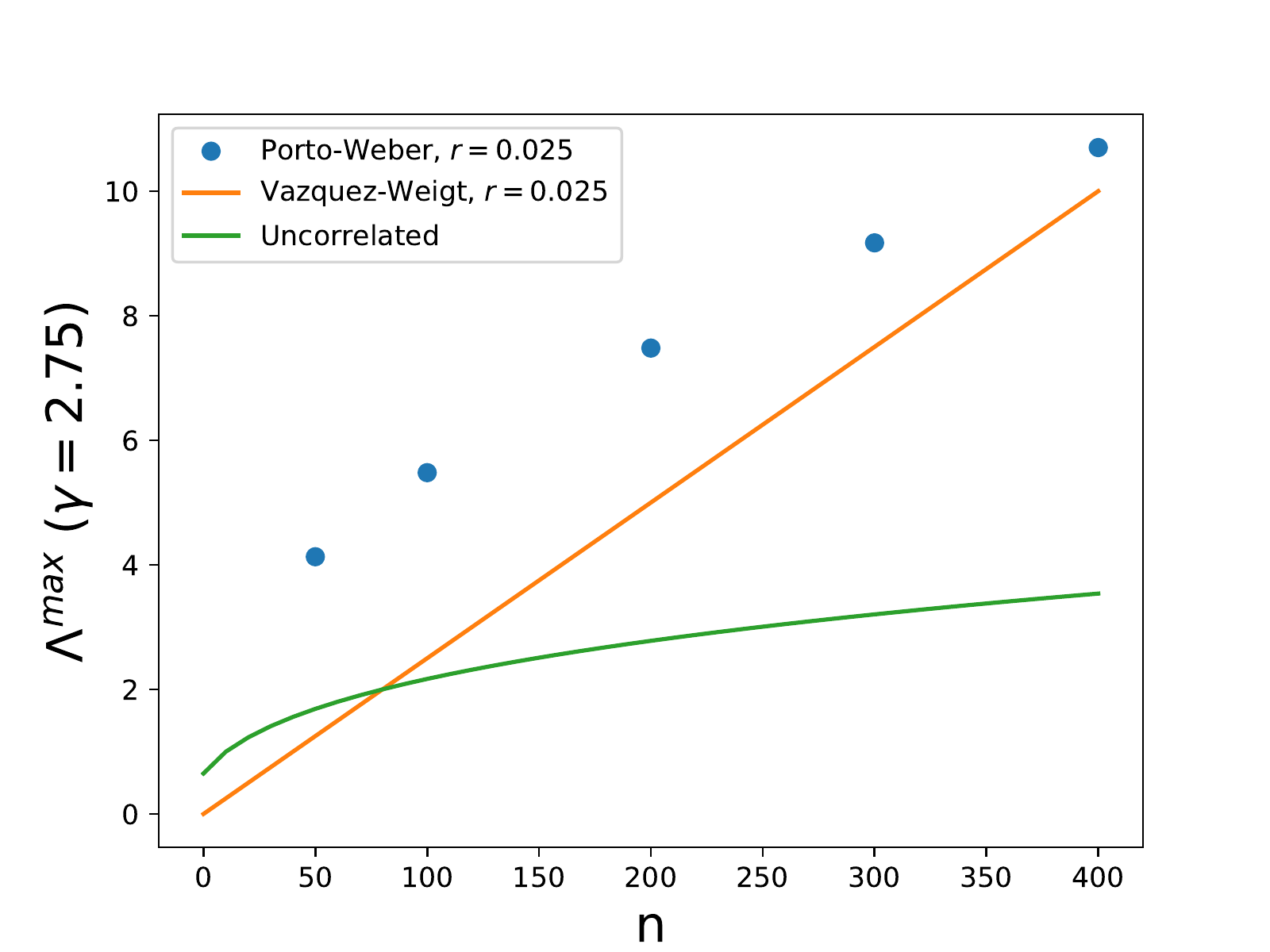}
\caption{Behavior, as a function of the maximum degree $n$, of the largest eigenvalue $\Lambda^{max}$ of the connectivity matrix, with scale-free exponent $\gamma=2.75$. Compare the case of $\gamma=2.5$ in Fig. \ref{La-max-25}.
} 
\label{La-max-275}
  \end{center}
\end{figure}

\begin{figure}[h]
  \begin{center}
\includegraphics[width=10.5cm,height=7.6cm]{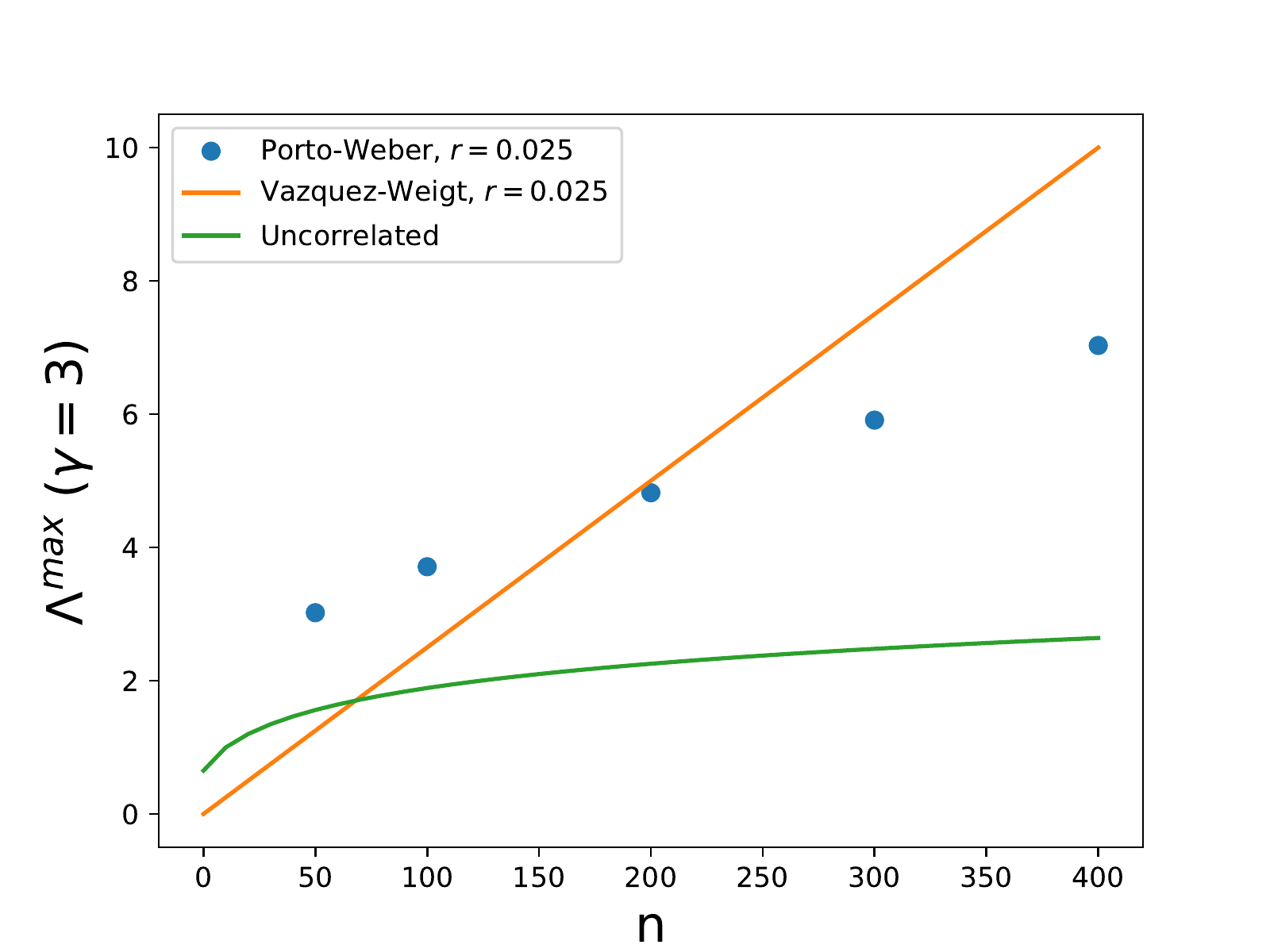}
\caption{Behavior, as a function of the maximum degree $n$, of the largest eigenvalue $\Lambda^{max}$ of the connectivity matrix, with scale-free exponent $\gamma=3$. Compare the cases of $\gamma=2.5$ in Fig. \ref{La-max-25} and $\gamma=2.75$ in Fig.\  \ref{La-max-275}.
} 
\label{La-max-3}
  \end{center}
\end{figure}

The connectivity matrix associated to $P^{Vaz-Wei}(h|k)$ is
\begin{equation}
    C^{Vaz-Wei}_{kh}=(1-r) \frac{khP(h)}{\langle k \rangle}+rk\delta_{kh} \, .
\end{equation}

In order to compute its eigenvalues $\Lambda^{(i)}$, we consider the determinant of the matrix $(C-\Lambda I)$, with elements
\begin{equation}
    C_{kh}^{Vaz-Wei}-\Lambda \delta_{kh}=(1-r) \frac{khP(h)}{\langle k \rangle} + (rk-\Lambda)\delta_{kh} \, .
\end{equation}
It is immediate to note that, when $(rk-\Lambda)=0$, the determinant of this matrix is zero because the matrix $khP(h)$ has rank 1. Thus we immediately find the $n$ eigenvalues
\begin{equation}
    \Lambda^{(i)}=ri \, , \qquad i=1,\ldots,n \, ,
\end{equation}
the largest of them being $\Lambda^{max}=rn$. When $n\to \infty$, this eigenvalue grows much faster than the largest eigenvalue for uncorrelated networks, especially if $\gamma$ approaches 3. For large networks, even a small value of $r$ like $r=0.025$ is sufficient for this to occur (see Figs.\ \ref{La-max-25}, \ref{La-max-275}, \ref{La-max-3}). 

Ref.\ \cite{vazquez2003resilience} reports the results of numerical simulations which, in retrospect, can be understood as being referred to a similar case of small $r$. On the analytical side, however, no general treatment was given, but only an approximated eigenvalue expansion valid for $r$ close to 1.

The conclusion is that for large scale-free networks the presence of a small diagonal assortative correlation guarantees a quick convergence to zero of the epidemic threshold. We recall that according to the criterion by Dorogovtsev and Mendez \cite{dorogovtsev2002evolution} the relation between network size $N$ (number of nodes) and maximum degree $n$ is $N \sim n^{\gamma-1}$. Therefore the behavior of $\Lambda^{max}$ is insensitive to $\gamma$ as a function of $n$ but not of $N$. Still, even for $\gamma=3$ the dependence of $n$ on $N$ is $n\sim \sqrt{N}$, which is fast compared to the very slow increase of $\Lambda^{max}$ in an uncorrelated network ($\Lambda^{max}\sim \ln N$).

\section{Examples of variations of a matrix $P(h|k)$ which do not modify its average nearest neighbour degree function $k_{nn}$.}
\label{4 Examples}

In this section we explore the more general question relative to multiplicity and concrete construction of variations of a correlation matrix $P(h|k)$ which do not modify the
average nearest neighbour degree function $k_{nn}$. If such a variation is represented as
\begin{equation}
P(h|k) \to P(h|k)+\Phi(h,k) \, ,
\label{variation}
\end{equation}
then the elements $\Phi(h,k)$ with $h, k=1, \ldots ,n$ are required to

(1) satisfy the Network Closure Condition (in this case, also the elements $P(h|k)+\Phi(h,k)$ do it, as one can easily check);

(2) satisfy the normalisation according to which $\sum_h (P(h|k)+\Phi(h,k))=1$
(for each $k$ the elements $P(h|k)+\Phi(h,k)$ give the probabilities that a vertex with degree k is connected to a vertex with degree $h$), which becomes in this case:
\[
\sum_h \Phi(h,k)=0, \ \ k=1, \ldots, n \, ;
\]

(3) leave $k_{nn}$ unchanged, which only occurs provided
\[
\sum_h h \Phi(h,k)=0, \ \ k=1, \ldots, n 
\]
holds true.

In addition, the inequalities
\begin{equation}
P(h|k) + \Phi(h,k) \ge 0  \ \quad \  \hbox{for all} \ h, k=1, \ldots, n
\label{ineq}
\end{equation}
must hold true.

To obtain condition (1), we start and assume, on the trail of Porto and Weber, 
\begin{equation}
\Phi(h,k)=\frac{h p_h}{\langle k \rangle} \, \phi_{h,k} \, ,
\label{PhiStamp}
\end{equation}
where $\phi_{h,k}$ is a symmetric matrix. Notice that from now on we will write
$p_h = P(h)$ for the sake of brevity.
Moreover, we will assume that each $p_h$ (for $h = 1, \ldots, n$) is different from zero.
The two conditions (2) and (3) then take the form
\begin{equation}
\left\{
\begin{array}{ccl}
	\sum_h h  p_h \phi_{h,k}&=&0 \ \quad \ \hbox{for all} \ k=1, \ldots, n     \\
	\sum_h h^2  p_h \phi_{h,k}&=&0  \ \quad \ \hbox{for all} \ k=1, \ldots ,n \, .
	\end{array}  \right.
	\label{2nx2neqs}
\end{equation}
We will first look for matrices $\phi = \{\phi_{h,k}\}$ which satisfy (7) and only afterwards,
in connection with some specific $P(h|k)$, will we check and specify when the inequalities (6) are satisfied too.

We narrow our search to the family of $n \times n$ symmetric matrices $\phi$, whose only nonzero elements are,
together with $\phi_{1,n}$ and $\phi_{n,1}$,
those
on the main diagonal, and those on the first diagonal below and on the first diagonal above the main diagonal:

\begin{equation}
\phi =
\left[
\begin{array}{cccccccccccc}
\phi_{1,1}  & \phi_{1,2}   & 0                 & 0 & 0 & \quad ... \quad & 0 & 0 & 0 & 0 & \phi_{1,n} \\ 
\phi_{1,2}  & \phi_{2,2}   & \phi_{2,3}    & 0 & 0 & \quad ... \quad & 0 & 0 & 0 & 0 & 0 \\
0               & \phi_{2,3}   & \phi_{3,3}    & \phi_{3,4} & 0 & \quad ... \quad & 0 & 0 & 0 & 0 & 0 \\
... & ... & ... & ... & ... & ... & ... & ... & ... & ... & ... \\
0 & 0 & 0 & 0 & 0 & \quad ... \quad &  0 & 0 & \phi_{n-1,n-2}     & \phi_{n-1,n-1}       & \phi_{n-1,n} \\
\phi_{1,n} & 0 & 0 & 0 & 0 & \quad ... \quad & 0 & 0 & 0           & \phi_{n-1,n}           & \phi_{n,n} \\    
\end{array}
\right] .
	\label{almtridPhi}
\end{equation}
For any such matrix $\phi$, solving system (7) amounts to solve a linear system of $2n$ equations in $2n$ variables (recall that $\phi$ is symmetric):
\begin{equation}
\left\{
\begin{array}{ccl}
	p_1 \phi_{1,1} + 2 \,p_2 \phi_{1,2} + n\, p_n \phi_{1,n} &=&0            \\
	p_1 \phi_{1,1} + 2\, p_2 \phi_{2,2} + 3\, p_3 \phi_{2,3} &=&0            \\
	2\, p_2 \phi_{2,3} + 3\, p_3 \phi_{3,3} + 4\, p_4 \phi_{3,4} &=&0            \\
	......... & \\
	(n-2)\, p_{n-2} \phi_{n-2,n-1} + (n-1)\, p_{n-1} \phi_{n-1,n-1} + n\, p_n \phi_{n-1,n} &=&0            \\
	p_1 \phi_{1,n} + (n-1) p_{n-1} \phi_{n-1,n} + n p_n \phi_{n,n} &=&0            \\
	p_1 \phi_{1,1} + 4\, p_2 \phi_{1,2} + n^2\, p_n \phi_{1,n} &=&0            \\
	p_1 \phi_{1,1} + 4\, p_2 \phi_{2,2} + 9\, p_3 \phi_{2,3} &=&0            \\
	4\, p_2 \phi_{2,3} + 9\, p_3 \phi_{3,3} + 16\, p_4 \phi_{3,4} &=&0            \\
	......... & \\
	(n-2)^2\, p_{n-2} \phi_{n-2,n-1} + (n-1)^2\, p_{n-1} \phi_{n-1,n-1} + n^2\, p_n \phi_{n-1,n} &=&0            \\
	p_1 \phi_{1,n} + (n-1)^2\, p_{n-1} \phi_{n-1,n} + n^2\, p_n \phi_{n,n} &=&0 \, .          \\
	\nonumber
	\end{array} \right.
	\label{2nx2nconcreteeqs}
\end{equation}
We rewrite this system as 
\begin{equation}
M \bf{x} = \bf{0} \, ,
\label{Mxiszero}
\end{equation}
where the matrix $M$ (also denoted by $M = M_n$) is given by 
$$
M =
\small{
\left[
\begin{array}{ccccccccccccccc}
p_{1}  & 0            & 0                 &  \quad ... \quad  & 0 &   0    &                   & 2 p_{2}             & 0           & 0       &  \quad ... \quad  & 0 &   n \, p_{n}    \\ 
0          & 2 \, p_{2}   & 0                 &  \quad ... \quad & 0 & 0 &                   & p_{1}   & 3 \, p_{3}       & 0            &  \quad ... \quad & 0 & 0  \\
0         & 0            & 3 \, p_{3}           & \quad ... \quad & 0 & 0 &                  & 0            & 2 \, p_{2}     & 4 \, p_4      & \quad ... \quad & 0 & 0\\
... & ... & ... & \quad ... \quad  & ... & ... & \ & ... & ... & ... & ... & ... & ... \\
0          & 0            & 0                &  \quad ... \quad  & (n-1) \, p_{n-1} & 0 &           & 0            & 0                &  \quad ... \quad  & ... & n  \, p_n & 0  \\ 
0          & 0   & 0                 &  \quad ... \quad  & 0 & n \, p_n &                    & 0 & 0                 &  \quad ... \quad  &  0   & (n-1) \, p_{n-1} & p_1  \\
p_{1}  & 0            & 0                 &  \quad ... \quad  & 0 &   0    &                    & 4 \, p_{2}             & 0     & 0        & 0     & 0 & n^2 \, p_{n} \\ 
0          & 4 \, p_{2}   & 0                 &  \quad ... \quad & 0 & 0 &                            & p_{1}   & 9 \, p_{3}       & 0           &  \quad ... \quad & 0 & 0 \\
0         & 0            & 9 \, p_{3}           & \quad ... \quad & 0 & 0 &                          & 0            & 4 \, p_{2}     & 16 \, p_4      & \quad ... \quad & 0 & 0\ \\
... & ... & ... & \quad ... \quad  & ... & ... & \ & ... & ... & ... & ... & ... & ... \\
0          & 0            & 0                &  \quad ... \quad  & (n-1)^2 \, p_{n-1} & 0 &           & 0            & 0                &  \quad ... \quad  & ... & n^2 \, p_{n}  & 0   \\ 
0          & 0   & 0                 &  \quad ... \quad  & 0 &  n^2 \, p_{n} &                         & 0   & 0             & 0     &  \quad ... \quad  &  (n-1)^2 \, p_{n-1} & p_1 \\
\end{array}
\right]} ,
$$
the unknown vector $\bf{x}$ is ordered as
${\bf{x}} = (\phi_{1,1}, \phi_{2,2},\phi_{3,3}, ..., \phi_{n,n}, \phi_{1,2}, \phi_{2,3},\phi_{3,4}, ... \phi_{n-1,n},\phi_{1,n})$ 
and 
$\bf{0}$ is the vector with all $2n$ components equal to zero.

It can be seen that, if each $p_h$ (for $h = 1, \ldots, n$) is different from zero, the matrix $M$ has determinant zero
whereas its rank is equal to $2n -1$.
Below, we first work out the calculations for the case $n=4$,
which is the smallest positive integer for which the particular structure of the matrix $\phi$ is clearly recognisable.
Then, we describe the procedure to handle the case with general $n$.

Denote by $M_4$ the matrix $M$ with $n=4$. 

\smallskip

{\bf{Proposition 1}}
The $8 \times 8$ matrix $M_4$ has determinant equal to zero and rank equal to seven.

\smallskip

{\itshape{Proof}}\! : 
The matrix $M_4$ has the form 
$$
M_4 =
\small{
\left[
\begin{array}{ccccccccccccccc}
p_{1}   & 0                 & 0                       &  0               & 2 \, p_{2}               & 0                   & 0                &   4 \, p_{4}    \\ 
0          & 2 \, p_{2}   & 0                        &  0                 & p_{1}                   & 3 \, p_{3}      & 0                 & 0  \\
0          & 0               & 3 \, p_{3}           &  0                  & 0                         & 2 \, p_{2}      & 4 \, p_4          & 0  \\
0          & 0                & 0                       &  4 \, p_{4}       & 0                      & 0                   & 3 \, p_{3}     & p_1\\
p_{1}   & 0                 & 0                       &  0             & 4 \, p_{2}             & 0                  & 0                &   16 \, p_{4}    \\ 
0          & 4\, p_{2}   & 0                        &  0                & p_{1}                   & 9 \, p_{3}      & 0                 & 0  \\
0          & 0               & 9 \, p_{3}           &  0                 & 0                         & 4 \, p_{2}      & 16 \, p_4       & 0  \\
0          & 0                & 0                       & 16 \, p_{4}           & 0                         & 0                   & 9 \, p_{3}    & p_1\\
\end{array}\right]} .
$$
By substituting the $5$-th row with that obtained as the difference of the $5$-th row minus the first row, we get a matrix,
whose determinant is easily seen to be equal to 
$$
\det (M_4) = p_1 \times \det (M_{4,\{7 \times 7\}}) \, ,
$$
where 
$$
M_{4,\{7 \times 7\}} =
\small{
\left[
\begin{array}{ccccccccccccccc}
2 \, p_{2}   & 0                        &  0                 & p_{1}                   & 3 \, p_{3}      & 0                 & 0  \\
0               & 3 \, p_{3}           &  0                  & 0                         & 2 \, p_{2}      & 4 \, p_4          & 0  \\
0                & 0                       &  4 \, p_{4}       & 0                      & 0                   & 3 \, p_{3}     & p_1\\
0                 & 0                       &  0             & 2 \, p_{2}             & 0                  & 0                &   12 \, p_{4}    \\ 
4\, p_{2}   & 0                        &  0                & p_{1}                   & 9 \, p_{3}      & 0                 & 0  \\
0               & 9 \, p_{3}           &  0                 & 0                         & 4 \, p_{2}      & 16 \, p_4       & 0  \\
0                & 0                       & 16 \, p_{4}           & 0                         & 0                   & 9 \, p_{3}    & p_1\\
\end{array}\right]} .
$$
By substituting the $5$-th row with that obtained as the difference of the $5$-th row minus $2$ times the first row, we get a matrix,
whose determinant is easily seen to be equal to
$$
\det (M_{4,\{7 \times 7\}}) = 2 p_2 \times \det (M_{4,\{6 \times 6\}}) \, ,
$$
where 
$$
M_{4,\{6 \times 6\}} =
\small{
\left[
\begin{array}{ccccccccccccccc}
3 \, p_{3}           &  0                  & 0                         & 2 \, p_{2}      & 4 \, p_4          & 0  \\
0                       &  4 \, p_{4}       & 0                      & 0                   & 3 \, p_{3}     & p_1\\
0                       &  0             & 2 \, p_{2}             & 0                  & 0                &   12 \, p_{4}    \\ 
0                        &  0                & - p_{1}                   & 3 \, p_{3}      & 0                 & 0  \\
9 \, p_{3}           &  0                 & 0                         & 4 \, p_{2}      & 16 \, p_4       & 0  \\
0                       & 16\, p_{4}           & 0                         & 0                   & 9 \, p_{3}    & p_1\\
\end{array}\right]} .
$$
With two similar further steps (i.e., iteratively suitably substituting the $5$-th row of a matrix) one easily finds that
\begin{equation}
\det (M_4) = p_1 \times 2 p_2 \times 3 p_3 \times 4 p_4 \times \det (M_{4,\{4 \times 4\}}) \, ,
\label{eqdetM4}
\end{equation}
where 
$$
M_{4,\{4 \times 4\}} =
\small{
\left[
\begin{array}{ccccccccccccccc}
2 \, p_{2}             & 0                  & 0                &   12 \, p_{4}    \\ 
 - p_{1}                   & 3 \, p_{3}      & 0                 & 0  \\
0                         & - 2 \, p_{2}      & 4\,  p_4       & 0  \\
0                         & 0                   & - 3 \, p_{3}    & - 3 \, p_1\\
\end{array}\right]} .
$$
The Laplace expansion of $\det (M_{4,\{4 \times 4\}})$, iteratively applied, gives
$$
2 \, p_2 \, [(- 3\, p_1) (3\, p_3 \, 4\, p_4)] + (-1) \, 12 \, p_4 \, [(- p_1) (- 2 \, p_2) (- 3 \, p_3)] = 0 \, .
$$
which in turn implies that $\det (M_4 )= 0$.

To conclude that the rank of $M_4$ is equal to seven it is sufficient, in view of (\ref{eqdetM4}), 
to prove that ${\rm rank} (M_{4,\{4 \times 4\}}) = 3$.
And this can be immediately seen, because for example the minor corresponding to the determinant of the triangular $3 \times 3$ matrix
$$
\small{
\left[
\begin{array}{ccccccccccccccc}
2 \, p_{2}             & 0                  & 0                    \\ 
 - p_{1}                   & 3 \, p_{3}      & 0                \\
0                         & - 2 \, p_{2}      & 4 \, p_4        \\
\end{array}\right]} 
$$
is different from zero (being $p_j \ne 0$ for all indices $j=1, \dots, n$ by assumption).
$\Box$

\smallskip

The proof strategy can be generalised for the case of the $2 n \times 2n$ matrix $M$ leading to the following result.

\medskip

{\bf{Proposition 2}}
For the $2n \times 2n$ matrix $M$ it is $\det (M) = 0$ and ${\rm rank} (M) = 2n - 1$.

\smallskip

{\itshape{Proof}}\! : 
By performing $n$ times a procedure similar to that in the proof of the previous proposition, namely 

- step $1$: substituting the $(n +1)$-th row of the matrix $M$ with that obtained as the difference of the $(n + 1)$-th row minus the first row, 
and then

-  step $2$: substituting the $(n +1)$-th row of the $(2n-1) \times (2n-1)$ matrix 
$M_{(2n-1) \times (2n-1)}$ obtained
after elimination of the first row and the first column 
from the matrix resulting from the previous step
with that obtained as the difference of the $(n + 1)$-th row minus $2$ times the first row, 
and then

- $...$

- step $n$: substituting the $(n +1)$-th row of the $(n +1) \times (n +1)$ matrix 
$M_{(n +1) \times (n +1)}$ obtained
after elimination of the first row and the first column 
from the matrix resulting from the previous step
with that obtained as the difference of the $(n + 1)$-th row minus $n$ times the first row, 

\noindent one finds that 
\begin{equation}
\det (M) = p_1 \times 2 \, p_2 \times ... \times n \, p_n \times \det (M_{\{n \times n\}}) \, ,
\label{eqdetM}
\end{equation}
where 
$$
M_{\{n \times n\}} =
\small{
\left[
\begin{array}{ccccccccccccccc}
2 \, p_{2}             & 0                  & 0                 &        \quad ... \quad                                        & 0       &         0                 &   (n^2 - n) \, p_{n}    \\ 
 - p_{1}                & 3\, p_{3}      & 0                  &        \quad ... \quad                                    & 0       &         0                 &        0       \\
0                         & - 2 \, p_{2}    & 4 \, p_4          &        \quad ... \quad                                     & 0       &         0                   &       0               \\
... & ... & ...                     &        \quad ... \quad                                      ... & ... & ... \\

 0        &         0                 &        0                    &        \quad ... \quad                             & (n - 1) \, p_{n - 1}                   &         0             &   0   \\ 
 0       &         0                 &        0                   &        \quad ... \quad                                      &  - (n - 2) \, p_{n - 2}                &        n \, p_{n}                &        0       \\
 0       &         0                 &        0                   &        \quad ... \quad                                     & 0     &       - (n - 1) \, p_{n - 1}  &        - (n - 1) \, p_{1}                \\
\end{array}\right]} .
$$
The determinant $\det (M_{\{n \times n\}})$ can be calculated by iteratively applying the Laplace expansion. It is not difficult to convince oneself that
\begin{equation}
\begin{array}{llll}
\det (M_{\{n \times n\}})  & = &   2 \, p_2 \, 3 \, p_3 \, ... \, n \, p_n \, \big ( - (n - 1)  \, p_1 \big )                                                                                                  & \  \\
                    \                & \  & + \, (- 1)^{n+1} \, (n^2 - n) \, p_n \, (- 1)^{n+1} \, \big (p_1 \, 2 \, p_2 \, ... \, (n-1) \, p_{n -1}\big )                                  & \  \\
                    \                & = &                \big (p_1 \, 2 \, p_2 \, ... \, (n-1) \, p_{n -1}\big ) \, \big ( - n^2 \, p_n + n \, p_n + n^2 \, p_n - n \, p_n \big )   & = 0 \, . \\
	\end{array}  
	\label{detMntimesn}
\end{equation}
Together, (\ref{eqdetM}) and (\ref{detMntimesn}) imply that $\det (M) = 0$.

It only remains to be proved that the rank of $M$ is equal to $2n - 1$. Also here, similarly as in the proof of Proposition $1$, we observe that
the $(n -1) \times (n -1)$ matrix obtained by deleting the last row and the last column in $M_{\{n \times n\}}$ has determinant equal to $p_1 \, 2 \, p_2 \, ... \, n \, p_n \ne 0$.
Hence, ${\rm rank} (M_{\{n \times n\}}) = n-1$ and, together with (\ref{eqdetM}), this completes the proof.
$\Box$

\medskip

Proposition $2$ implies that the eigenspace of $M$ is one-dimensional and this in turn means that 
the equation (\ref{Mxiszero}) admits infinitely many solutions; precisely, there is a one-parameter family of them.
By way of example, let us consider the following low-dimensional case.

\medskip

{\bf{Example 1}}
Let $n = 4$ and let a Markovian network $\mathcal{N}_4$
with degree distribution $P(h)$ and correlation matrix $P(h|k)$ be given. 
Assume that $p_h = P(h) \ne 0$ for $h = 1, \ldots, n$.
The function $k_{nn}$ pertaining to $\mathcal{N}_4$
is also the average nearest neighbour degree 
of the networks which have the same degree distribution $p_h$ as $\mathcal{N}_4$ and correlation matrix of the form $P(h|k)+\frac{h p_h}{\langle k \rangle} \phi_{h,k}$,
the only nonzero elements of the symmetric matrix $\phi_{h,k}$ being
\begin{equation}
\begin{array}{llll}
\phi_{1,1} \, ,\quad  & \phi_{2,2} =  \frac{3 \, p_1^2}{4 \, p_2^2} \, \phi_{1,1}  \, ,\quad & 
\phi_{3,3}  = \frac{p_1^2}{3 \, p_3^2} \, \phi_{1,1}  \, ,\quad & \phi_{4,4}  = \frac{ p_1^2}{16 \, p_4^2} \, \phi_{1,1} \, , \\[1.5ex]
\phi_{1,2} = - \frac{3 \, p_1}{4 \, p_2} \, \phi_{1,1} \, ,\quad & \phi_{2,3} =  - \frac{ p_1^2}{4 \, p_2 \, p_3} \, \phi_{1,1} \, ,\quad & 
\phi_{3,4}  = - \frac{ p_1^2}{8 \, p_3 \, p_4} \, \phi_{1,1} \, ,\quad & \phi_{1,4}  = \frac{p_1}{8 \, p_4} \, \phi_{1,1} \, , \\
	\end{array}  
	\label{onepf4}
\end{equation}
all of them expressed in terms of a unique parameter $\phi_{1,1}$,
together with those symmetrically positioned with respect to the main diagonal,
provided the inequalities
\begin{equation}
\phi_{h,k} \ge - \, \frac{\langle k \rangle}{h p_h}  \, P(h|k) \ \quad \  \hbox{for all} \ h, k=1, \ldots, n
\label{ineqcase4}
\end{equation}
are satisfied.

We recall here that $P(h|k) \ge 0$ for all $h, k=1 \ldots n$. Therefore, the inequalities to be checked are in fact those relative to the nonzero elements $\phi_{h,k}$.
In this example ($n = 4$), they can be expressed as
\begin{equation}
\left\{
\begin{array}{cclll}
	\phi_{h,h}   & \ge & - \, \frac{\langle j \rangle}{h p_h}  \, P(h|h) \ \ \ \hbox{for all} \  h =1, \ldots, 4     \\ 
	\phi_{1,2}   & \ge & \max \Big\{- \, \frac{\langle j \rangle}{p_1}  \, P(1|2)      , \, - \, \frac{\langle j \rangle}{2 p_2}  \, P(2|1)  \Big \} \ &    \  \\
	\phi_{2,3}   & \ge & \max \Big\{- \, \frac{\langle j \rangle}{2 p_2}  \, P(2|3)    , \, - \, \frac{\langle j \rangle}{3 p_3}  \, P(3|2)  \Big \} \ &    \  \\
        \phi_{3,4}   & \ge & \max \Big\{- \, \frac{\langle j \rangle}{3 p_3}  \, P(3|4)     , \, - \, \frac{\langle j \rangle}{4 p_4}  \, P(4|3)  \Big \} \ &    \  \\
        \phi_{1,4}   & \ge & \max \Big\{- \, \frac{\langle j \rangle}{p_1}  \, P(1|4)       , \, - \, \frac{\langle j \rangle}{4 p_4}  \, P(4|1)  \Big \} \ &   \  \, . 
        	\end{array}  \right.
	\label{ineqconcrete}
\end{equation}

Explicit treatment of an example requires fixing both the values of the elements of the degree distribution $p_h$ and correlation matrix $P(h|k)$.
We here consider three cases, each of them relative to a scale-free network with $p_h = c/h^{\gamma}$, where $\gamma \in (2,3)$, and $P(h|k)$
constructed according to the following algorithms (see \cite{bertotti2019evaluation,bertotti2021comparison}):

\smallskip

$\bullet$ Case $1$:
Let
${P_0}(h|k) = |h - k|^{ - 1 }$ if $h < k$, ${P_0}(h|k) = 1$ if $h = k$
and 
\begin{equation}
P_0(h|k) = P_0(k|h)\frac{{{h^{1 - \gamma }}}}{{{k^{1 - \gamma }}}} \quad {\rm if} \ h > k \, .
\label{Pnonsym}
\end{equation}
Call
$C_k = \sum_{h = 1}^n P_0(h|k)$ for any $k = 1,...,n$
and let $C_{max} = \max_{k=1,...,n} \, C_k$.
Then, re-define the correlation matrix
by setting the elements on the diagonal equal to
\begin{equation*}
{P_1}(k|k)= C_{max} - C_k, \quad k = 1,...,n \, ,
\label{Pprimefirst}
\end{equation*}
and leaving the other elements unchanged: ${P_1}(h|k) = {P_0}(h|k)$ for $h\ne k$.
Finally, normalize the entire matrix by setting
\begin{equation*}
{P}(h|k) = \frac{1}{(C_{max} - 1)} \,  {P_1}(h|k), \quad h,k = 1,...,n \, .
\label{P_Dsecondfirst}
\end{equation*}

$\bullet$ Case $2$:
Let
${P_0}(h|k) = 1 -  \frac{1}{N} \, |h - k|$ if $h \le k$
and ${P_0}(h|k)$ with $h > k$ as in (\ref{Pnonsym}). Then, proceed as in the previous case
to get elements $P(h|k)$ which satisfy the normalisation $\sum_{h = 1}^n P(h|k) =1$.

$\bullet$  Case $3$:
Let
${P_0}(h|k) = \displaystyle{e^{- \, \frac{(h - k)^2}{n^2}}}$ if $h \le k$
and ${P_0}(h|k)$ with $h > k$ as in (\ref{Pnonsym}). Again, proceed as above
to get elements $P(h|k)$ which satisfy the normalisation $\sum_{h = 1}^n P(h|k) =1$.

\smallskip

Straightforward calculations (performed with Mathematica) yield for example that the inequalities (\ref{ineqconcrete}) are satisfied

\smallskip

- in Case $1$, if $\gamma = 2.1$, provided $0 \le \phi_{1,1} \le 0.12908$;

- in Case $1$, if $\gamma = 2.5$, provided $0 \le \phi_{1,1} \le 0.06478$;

- in Case $1$, if $\gamma = 2.9$, provided $0 \le \phi_{1,1} \le 0.03369$;

\smallskip

- in Case $2$, if $\gamma = 2.1$, provided $0 \le \phi_{1,1} \le 0.12010$;

- in Case $2$, if $\gamma = 2.5$, provided $0 \le \phi_{1,1} \le 0.06012$;

- in Case $2$, if $\gamma = 2.9$, provided $0 \le \phi_{1,1} \le 0.03119$;

\smallskip

- in Case $3$, if $\gamma = 2.1$, provided $0 \le \phi_{1,1} \le 0.11247$;

- in Case $3$, if $\gamma = 2.5$, provided $0 \le \phi_{1,1} \le 0.05618$;

- in Case $3$, if $\gamma = 2.9$, provided $0 \le \phi_{1,1} \le 0.02876$.

\smallskip

When calculating the connectivity matrix $C_{k,h}$ in correspondence to correlations $P(h|k)$ and then also 
in correspondence to correlations $P(h|k)+\Phi(h,k)$ in the Cases $1, 2, 3$ above,
for $\gamma = 2.1, 2.5, 2.9$ and (various) values of $\phi_{1,1}$ compatible with the intervals just found,
one observes what follows.
In passing from $P(h|k)$ to $P(h|k)+\Phi(h,k)$ (namely, by taking an admissibile positive $\phi_{1,1}$ rather than $\phi_{1,1}=0$),
in all Cases, $1$, $2$ and $3$, the largest eigenvalue of $C_{k,h}$ increases (and, accordingly, the epidemic threshold decreases).  

In contrast, if one takes an uncorrelated network, by this meaning a network for which

$\bullet$ Case $4$:
$P(h|k) = h P(h) / \langle k \rangle$, 

and considers then
the matrix $P(h|k)+\Phi(h,k)$ with elements $\Phi(h,k)$ constructed according to (\ref{PhiStamp}) and (\ref{almtridPhi}),
one notices the following fact: the largest eigenvalue of the connectivity matrix $C_{k,h}$
remains equal to $\frac{\langle k^2 \rangle}{\langle k \rangle}$ when the parameter $\phi_{1,1}$ varies in the interval which guarantees the meaningfulness of the variation
($-0.01234 < \phi_{1,1} < 0.01603$).
A subtle and interesting situation is taking place: on one hand,
for networks constructed considering the elements $P(h|k)+\Phi(h,k)$ as done here 
it is no more true that
the conditional probability that a vertex of degree $k$ is connected to a vertex of degree $h$ is independent of $k$;
on the other hand, the average nearest neighbor degree function $k_{nn}(k)$ of these networks is the same of 
that of a strictly uncorrelated network; it is constant (and equal to $\frac{\langle k^2 \rangle}{\langle k \rangle}$, coinciding with the $k_{nn}(k)$ of the 
original uncorrelated network).
$\Box$

\medskip

{\bf{Remark 1}}
Beside the choice of taking
symmetric matrices $\phi$ as in (\ref{almtridPhi}),
other choices can be performed. 
They lead both to cases in which the largest eigenvalue of the connectivity matrix in correspondence to the matrix $P(h|k)+\Phi(h,k)$
is greater than the one obtained in correspondence to the matrix $P(h|k)$
as to cases in which this eigenvalue is smaller. 
$\Box$

In any case, one can conclude that networks with the same $k_{nn}$, but different correlation matrices
can exhibit different epidemic thresholds.

\bigskip
\bigskip
\bigskip

\bibliographystyle{unsrt}
\bibliography{refs-pwa}

\end{document}